\documentclass[preprint,aps,prx,superscriptaddress,showpacs,nofootinbib,citeautoscript,floatfix]{revtex4-2}

\usepackage{graphicx}
\usepackage{dcolumn}
\usepackage{bm}
\usepackage{color}
\usepackage[utf8]{inputenc}
\usepackage[T1]{fontenc}
\usepackage{mathptmx}
\usepackage{etoolbox}
\usepackage{booktabs}
\usepackage{multirow}
\usepackage{textgreek}
\usepackage{graphicx}
\usepackage{dcolumn}
\usepackage{bm}
\usepackage{flafter}
\usepackage{float}
\usepackage{lettrine}
\usepackage{amsmath}
\usepackage{amssymb}
\usepackage{color}
\usepackage{epsfig}
\usepackage{fix-cm}
\usepackage{braket}
\usepackage{textalpha}
\makeatother
\begin{document}

\title{On the ambient conditions crystal structure of AgSbTe$_2$}

\author{Baihong Sun}
\affiliation {Materials Science and Engineering Department, Guangdong Technion-Israel Institute of Technology, Shantou 515063, China}
\affiliation{Department of Materials Science and Engineering, Technion-Israel Institute of Technology, Haifa 3200003, Israel}
\author{Sergei Grazhdannikov}
\affiliation{Department of Materials Science and Engineering, Technion-Israel Institute of Technology, Haifa 3200003, Israel}
\author{Muhamed Dawod}
\affiliation{Department of Materials Science and Engineering, Technion-Israel Institute of Technology, Haifa 3200003, Israel}
\author{Lunhua He }
\affiliation{Beijing National Laboratory for Condensed Matter Physics, Institute of Physics, Chinese Academy of Sciences, Beijing 100190, PR China}
\affiliation{Spallation Neutron Source Science Center, Dongguan 523803, PR China}
\affiliation{Songshan Lake Materials Laboratory, Dongguan 523808, PR China}
\author{Jiazheng Hao }
\affiliation{Spallation Neutron Source Science Center, Dongguan 523803, PR China}
\affiliation{Institute of High Energy Physics, Chinese Academy of Sciences, Beijing, 100049, PR China}
\author{Thomas Meier}
\email{thomas.meier@hpstar.ac.cn}
\affiliation{Shanghai Key Laboratory MFree, Institute for Shanghai Advanced Research in Physical Sciences, Pudong, Shanghai, 201203, China} 
\affiliation{Center for High Pressure Science and Technology Advanced Research (HPSTAR), Beijing 100193, China.}
\author{Yansun Yao}
\email{yansun.yao@usask.ca }
\affiliation{Department of Physics and Engineering Physics, University of Saskatchewan Saskatoon, Saskatchewan S7N 5E2, Canada  }
\author{Yaron Amouyal}
\email{amouyal@technion.ac.il}
\affiliation{Department of Materials Science and Engineering, Technion-Israel Institute of Technology, Haifa 3200003, Israel}
\author{Elissaios Stavrou}
\email{elissaios.stavrou@gtiit.edu.cn}
\affiliation {Materials Science and Engineering Department, Guangdong Technion-Israel Institute of Technology, Shantou 515063, China}
\affiliation{Department of Materials Science and Engineering, Technion-Israel Institute of Technology, Haifa 3200003, Israel}
\affiliation{Guangdong Provincial Key Laboratory of Materials and Technologies for Energy Conversion, Guangdong Technion-Israel Institute of Technology, Shantou 515063, China}

\date{\today}%

\begin{abstract}
We present a combined X-ray and neutron diffraction, Raman spectroscopy, and $^{121}$Sb NMR studies of AgSbTe$_2$, supported by first-principles calculations aiming to elucidate its crystal structure. While diffraction methods cannot unambiguously resolve the structure, Raman and NMR data, together with electric field gradient calculations, strongly support the rhombohedral $R\overline{3}m$ phase. Moreover, the agreement between experimental and calculated Raman spectra further corroborates this result, resolving  the 60-year sold debate about the exact crystal structure of the AgSbTe$_2$ compound.

\end{abstract}

\maketitle

The global growing demands for energy consumption and the accompanying environmental issues have forced the scientific community  to find solutions to overcome the energy crisis and the development of sustainable and environmentally friendly energy resources has become critical \cite{shi2020advanced}. In this context, the application of thermoelectric (TE) materials that can directly convert energy from heat to electricity and vice versa, is a very promising way to overcome present challenges.  AgSbTe$_2$, a ternary chalcogenide TE material, appears promising mainly due to its very low thermal conductivity 0.5–0.7 Wm$^{-1}$K$^{-1}$ \cite{morelli2008,carlton2014,hoang2007} and its very narrow band gap of 7.6 meV \cite{jovovic2008}. Recently, it has gained significant attention for achieving a high figure of merit (ZT) of approximately 2.6 at 573 K through cadmium doping \cite{Roychowdhury2021}.

However, its crystal  structure at room temperature (RT), which is of fundamental importance for understanding and tuning its corresponding TE properties,  has been the subject of a long debate. Initially, it was considered to have a cubic $Fm\overline{3}m$ disordered NaCl-type partial solid solution structure\cite{geller1959}. In this crystal structure,  Ag and Sb occupy the 4a Wyckoff positions (Na-site) with 50\% occupancy.  This solid solution arrangement was challenged by Quarez $et$ {} $al.$ \cite{quarez2005} and three ordered structures, with space groups $Pm\overline{3}m$, $P4/mmm$ and  $R\overline{3}m$ ($\alpha$-NaFeO$_2$-type), have been proposed as the ones that can best index the corresponding single-crystal X-ray diffraction (XRD) results.

Subsequent first-principles calculations \cite{hoang2007} suggested that the ordered $Fd\overline{3}m$ (D4) crystal structure has lower energy compared to the structures identified by Quarez $et$ {} $al.$ However, in the same study, rhombohedral $R\overline{3}m$ (L$1_{1}$)  was found to have practically degenerate energy with the D4 structure. Indeed, the negligible energy difference between the D4 and L$1_{1}$ structures was confirmed by later first-principles calculations \cite{barabash2008,Rezaei2014,Szczypka2018}. Further first-principles studies of the temperature-dependent Helmholz free energies of the cubic, tetragonal, and rhombohedral polymorphs indicate slight differences between these three values around RT, so that evident identification of the stable structure remains unequivocal \cite{Amouyal2013, Amouyal2014,Amouyal2016}. Therefore, the exact crystal structure of AgSbTe$_2$ remains an open question.  Accurate determination of  its  crystal structure  is of critical importance for relevant theoretical calculations aiming to elucidate and tune its thermal and electrical transport properties followed by its high TE performance.

In this study, our goal is to accurately determine the  crystal structure of AgSbTe$_2$ based on the previously proposed crystal structures $Fm\overline{3}m$, $Pm\overline{3}m$, $P4/mmm$, $L1_{1}$ and $D4$. We used concomitant  XRD, neutron diffraction (ND), Raman spectroscopy and solid-state nuclear magnetic resonance (NMR)  experimental probes combined with first-principle density functional theory (DFT) calculations of the corresponding Raman spectra. Both diffraction techniques   are unable to provide a definitive structural determination. On the other hand, Raman as well as NMR spectroscopy results provide a definitive identification of the rhombohedral L$1_{1}$ crystal structure.  


AgSbTe$_2$ specimens  were synthesized from high-purity raw materials, including silver (Ag, 99.999\%, Alfa Aesar) granules, antimony (Sb, 99.999\%, Alfa Aesar) granules and tellurium (Te, 99.99\%, Strem Chemicals) broken ingots. These raw materials were mixed in the appropriate molar ratios to yield the desired composition resulting in a total ingot mass of 20 g. The mixture was loaded into quartz ampules, which were evacuated to 5 × 10$^{-5}$ torr residual pressure and sealed. The sealed ampules were annealed to 1000 $^o$C in a programmable vertical tube furnace, where their molten content was mixed three times during a 32-h soaking period. Afterwards, the ampules were cooled by quenching in ice-water to obtain AgSbTe$_2$ ingots, which were further ground into fine powders sieved with a  56 $\mu$m mesh.

The microstructure and compositional characterization of the samples was carried out using a Zeiss® Ultra Plus high-resolution scanning electron microscope (HR-SEM), equipped with a Schottky field-emission electron gun, as well as the FEI Quanta 200 SEM equipped with a 10 mm$^2$ active area Oxford® INCA X-sight EDS detector with an energy resolution of 133 eV. The elemental composition was further examined using energy dispersive X-ray spectroscopy (EDX) measurements with a 8 kV electron beam and the results (23.4±0.5 at\% of Ag, 26.8±0.3 at\% of Sb and 49.8±0.6 at\% of Te) confirmed the synthesis of AgSbTe$_2$.

Powder XRD measurements were performed with the Cu K$\alpha$$_1$ ($\lambda$=1.5406\r{A}) X-ray line (Rigaku MiniFlex). Powder ND was performed using the general purpose powder diffractometer (GPPD) located at the China Spallation Neutron Source (CSNS). The GPPD is a time-of-flight (TOF) diffractometer with a neutron bandwidth of 4.8 \AA, providing a maximum resolution of $\Delta$d/d=0.15\%. The neutron pattern data in this study were acquired from three different banks of GPPD: 150$^0$, 90$^0$, and 30$^0$, corresponding to the central solid angles of the detector. The d-space ranges were between 0.05-2.7\AA,  0.06-4.3\AA, and   0.12–28.11\AA, respectively \cite{He2023,Hao2023}. The sample under investigation was loaded into 9 mm diameter TiZr can and all measurements were performed at RT. 

Raman studies were performed using a custom-made confocal micro-Raman system with the 660 nm line of a solid-state laser for excitation in back scattering geometry. The laser probing spot dimension was $\approx$ 4 $\mu$m. Raman spectra were recorded with a spectral resolution of 2 cm$^{-1}$ using a single-stage grating spectrograph equipped with a charge-coupled device (CCD) array detector. The laser power on specimens was kept below 0.2 mW, to avoid any laser-induced decomposition. Ultra-low-frequency solid-state notch filters allowed us to measure Raman spectra down to 10 cm$^{-1}$ \cite{Hinton2019}. 


NMR spectroscopy was conducted at 5.87 T ($^1$H frequency 250 MHz) using homebuilt probes. $^{121}$Sb NMR powder spectra were recorded over ±2 MHz around 59.83 MHz using frequency-stepped Hahn echoes\cite{Hahn1950} with Gaussian amplitude modulation. The transmitter frequency was incremented in 25 kHz steps, employing 70~$\mu$s pulses (40 kHz excitation bandwidth) at 0.4 W pulse power. With a 100 ms repetition time, the average pulse power was under 0.4 mW to minimize sample decomposition. Due to rapid spin coherence decay ($T_2^* < 5\,\mu$s), echo pulse separations were kept below 6 µs. The full spectrum was reconstructed from individual echo intensities versus carrier frequency.

Electric field gradient calculations followed standard definitions of quadrupolar parameters\cite{Man2006, Haase1995}. Atomic positions near Sb nuclei were iterated for $Fm\overline{3}m$ and $R\overline{3}m$ structures with a 50 ~\AA {} cutoff. On-the-fly checks ensured convergence of Cq and η. Spectral simulations of full quadrupolar powder patterns employed the contour analysis method of Hughes and Harris \cite{Hughes2016}, accounting for first- and second-order quadrupolar interactions. The line shape was convoluted with a Voigt profile (40 kHz Lorentzian and Gaussian components) to model natural linewidth effects.


Raman frequencies and intensities were calculated from first principles using the linear response pseudopotential plane-wave approach within the framework of density functional perturbation theory \cite{baroni2001}. Local density approximation \cite{perdew1992} was applied, and Troullier-Martins norm-conserving pseudopotentials \cite{troullier1991} were employed, using valence states of 4d$^{10}$5s$^1$, 5s$^2$5p$^3$, 5s$^2$5p$^4$ for Ag, Sb, and Te, respectively. Integration over the first Brillouin zone was performed using a 16×16×16 Monkhorst-Pack mesh \cite{monkhorst1976}. All calculations were carried out with the QUANTUM-ESPRESSO package \cite{giannozzi2009,giannozzi2017}. The basic concept is solving the change of charge polarizations due to the perturbations introduced by the vibration (zone-center phonon). The background theory and computational techniques have been described elsewhere \cite{baroni2001}.

The experimental XRD pattern compared with the calculated patterns, using the POWDER CELL program \cite{Kraus1996} for the corresponding previously suggested crystal structures,   is shown in Fig. 1. It is clear that  powder XRD results cannot provide a conclusive answer to the exact crystal structure, in agreement with previous studies, since it can be fitted with any of the suggested crystal  structures. The additional observed Bragg   peaks, besides those associated to  AgSbTe$_2$,  in the experimental XRD pattern can be assigned to the $\beta$-Ag$_2$Te phase, see inset of Fig. 1. 
The precipitation of $\beta$-Ag$_2$Te was also observed in previous studies \cite{hu2025,kim2024}. 

\begin{figure}[h]
\centering
\includegraphics[width=\linewidth]{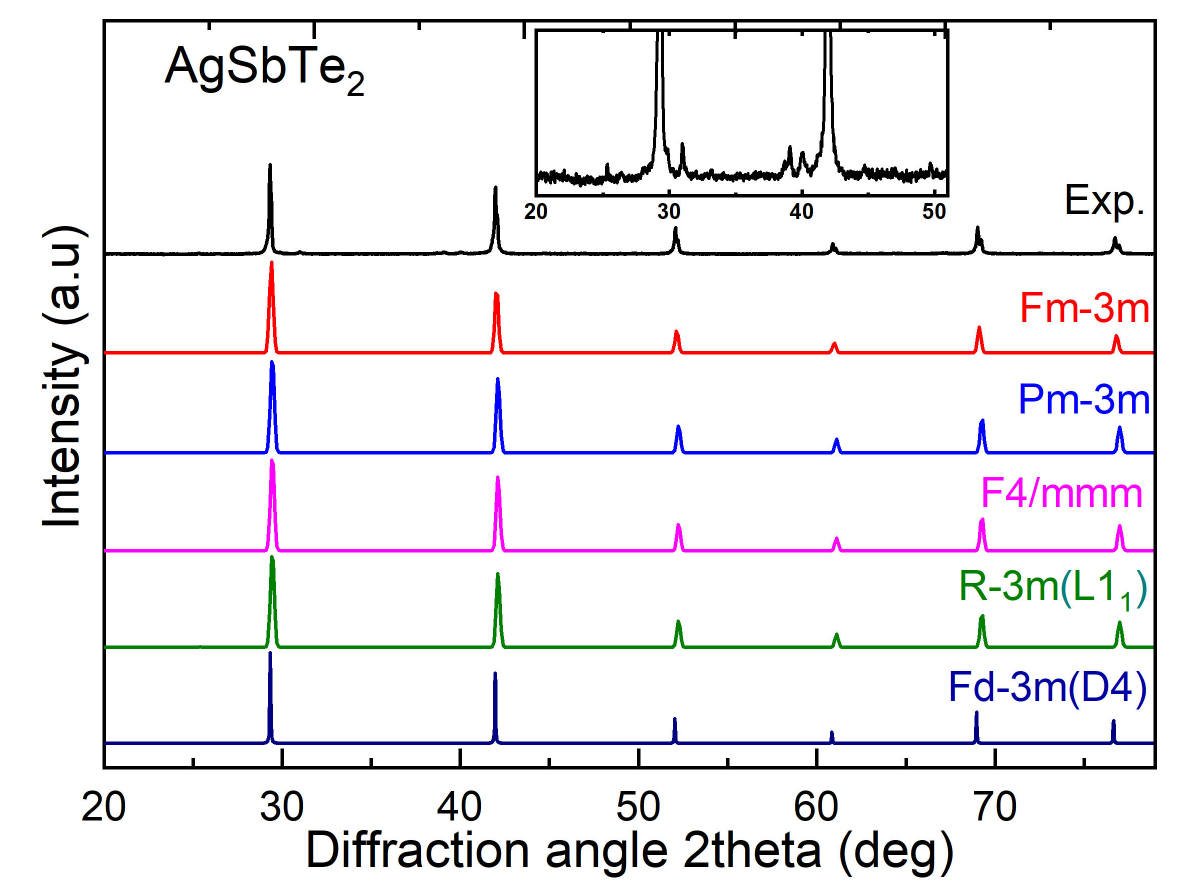}
\caption{Experimental XRD pattern of AgSbTe$_2$ (black) in comparison with the calculated patterns of the previous proposed crystal structures. The inset showcase the Bragg peaks of the Ag$_2$Te impurity phase.}
\end{figure}

Rietveld refinements, using the $GSAS-II$ software \cite{Toby2013}, were performed on the ND pattern, using the previously proposed  crystal structures.  The  $Fm\overline{3}m$ structure was  used in order to determine the volume fraction of the $\beta$-Ag$_2$Te, as shown in Fig. S1(a) \cite{supp}. The volume fractions of the AgSbTe$_2$ and the $\beta$-Ag$_2$Te phases were determined to be 95.81\% and 4.191\%, respectively. Subsequently, the Goodness of Fit (GOF), the  reduced  $\chi^2$ and R$_w$ parameters for the various examined crystal structures were determined, see Figs. 2 and S1. These parameters were used as the criteria for assessing which structure provides the better fit. The corresponding results  for the  $Fm\overline{3}m$, $Fd\overline{3}m$ (D4), $R\overline{3}m$ (L1$_1$), $Pm\overline{3}m$ and $P4/mmm$ crystal structures are summarized in Table I. As in the case of powder XRD, powder ND results cannot   provide a definitive identification of the crystal structure of AgSbTe$_2$. In fact, Rietveld refinement calculations for different preselected structures indicate similar results; see Table I.  The  shortcoming of ND is most probably associated with the similar coherent neutron scattering cross sections of Ag (4.407 b), Sb (3.9 b) and Te (4.23 b) \cite{Sears1992}.

\begin{table}[h]
\centering
\footnotesize
 \caption{Rietveld refinement results of ND patterns of AgSbTe$_2$ for the various previously suggested crystal structures.  The  Fm-3m+$\beta$-Ag$_2$Te mixture was independently used in one of the refinements, aiming to  determine the volume fraction of secondary phase $\beta$-Ag$_2$Te}
 \begin{ruledtabular}
    \begin{tabular}{ccccccccc} 
    Space Group& a [\AA]&  c [\AA]& Ag& Sb& Te& GOF&  R-$\chi^2$&  R$_w$  \\ \hline
    Fm-3m+$\beta$-Ag$_2$Te& 6.077 &  -& 4a& 4a& 4b& 2.13&  4.54& 3.07\%\\
    Fm-3m&                  6.077& -& 4a& 4a& 4b& 3.42&  11.72& 5.08\% \\ 
    Fd-3m (D4)&             12.154&  -& 16c& 16d& 32e& 3.06& 9.38&  4.52\% \\
    R-3m (L1$_1$)&          4.295&  21.073& 3b& 3a& 6c& 3.00&  8.97&  4.42\% \\
    Pm-3m&          6.077& -& 3c& 1b, 3c& 1a, 3d& 3.10&  9.64&  4.59\% \\ 
    P4/mmm&                 4.297&  6.077& 1c& 1b& 1a, 1d& 3.11& 9.68&  4.61\%  \\

\end{tabular}
\end{ruledtabular}
\end{table}

\begin{figure}[h]
\centering
\includegraphics[width=\linewidth]{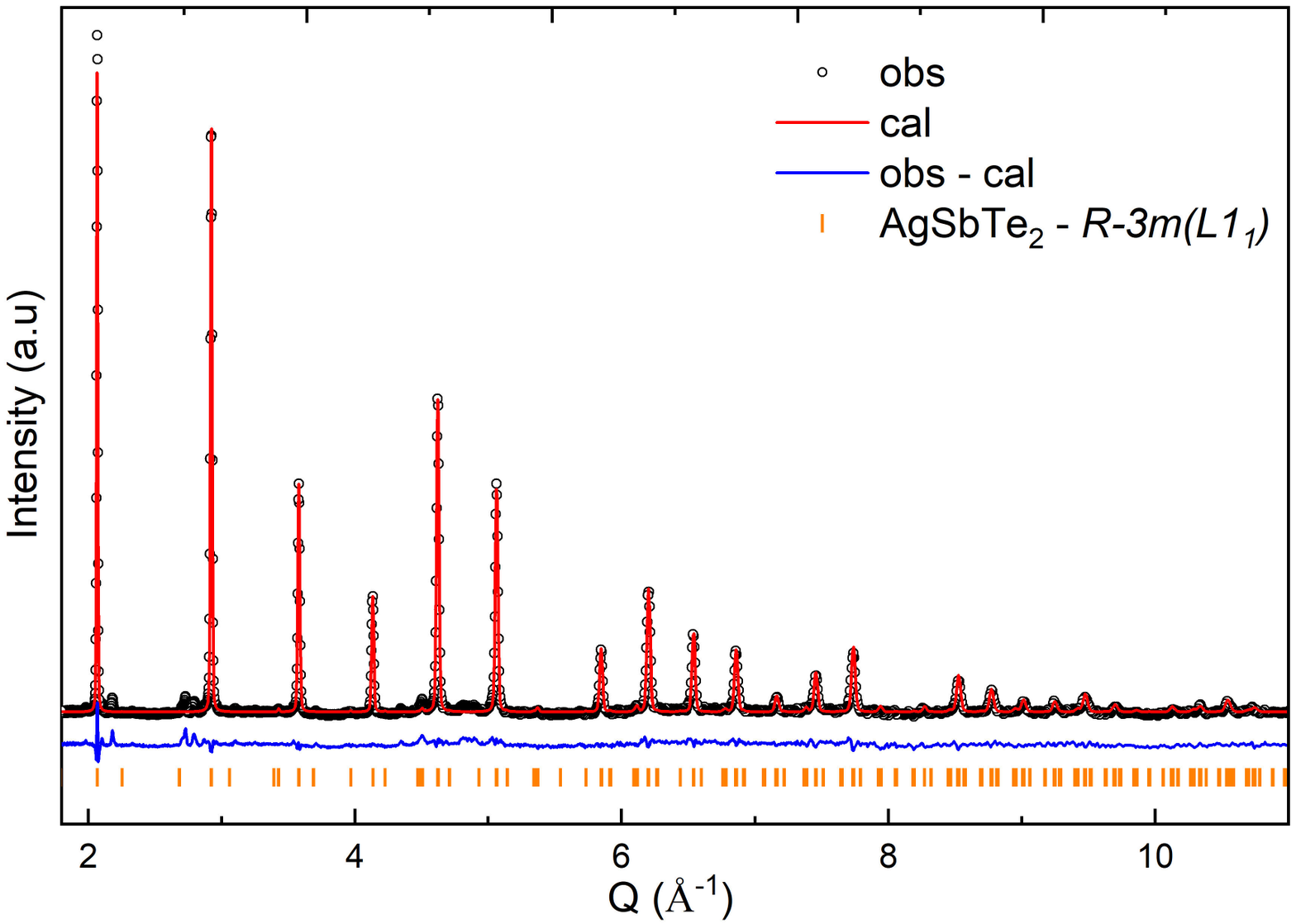}
\caption{Rietveld refinement of the experimental ND  pattern of AgSbTe$_2$ using the \textit{R-3m}  structural models.}
\end{figure}

According to group theory, the Raman active zone center modes for the L1$_1$ crystal structure  are $\widetilde{\Gamma}_{R}$ = $A_{1g}+E_g$ \cite{Kroumova2003}.  Both modes are associated with the Te (6c) atoms vibrations. This crystal structure is characterized by alternating Ag-Te-Sb-Te layers,  perpendicular to the c-axis. For the $A_{1g}$ mode, two adjacent Te layers vibrate  rigidly against each other and parallel to the c-axis, while for the $E_g$ mode, two adjacent Te layers vibrate against each other and perpendicular to the c-axis \cite{Brueesch1971, Yelisseyev2015}, see Fig. S2. In contrast, all the other previously suggested/proposed crystal structures are Raman inactive.

Although the overall Raman intensity is low and the relevant Raman peaks are close to the Rayleigh line, the Raman spectrum of this study, see Fig. 3, clearly indicates the existence of two Raman active modes at 114.7 and 149.1 cm$^{-1}$. Furthermore, the relative intensities and frequencies are in fair agreement with those calculated for the L1$_1$ crystal structure of AgSbTe$_2$. The agreement becomes even better after the subtraction of a simple baseline, reflecting the proximity to the Rayleigh line, see blue spectrum in Fig. 3. We attribute the apparent high-intensity of the Rayleigh to the, effectively, metallic character of AgSbTe$_2$. Indeed, previous studies document that AgSbTe$_2$  is either a semimetal \cite{Hoang2016} or a very narrow band gap ($E_g$ $\approx$ 7.6 meV, below the thermal energy at ambient conditions) semiconductor with  a high  number of  thermally excited  and high-mobility electrons \cite{jovovic2008}. In addition to the agreement between the experimental and calculated spectra, further evidence support the assignment of the observed Raman modes originating from the$R\overline{3}m$ structure. Indeed, the  overall Raman spectrum  ``shape'' and  mode frequencies fits well with those  of other $\alpha$-NaFeO$_2$-type structure compounds, such as LiNiO$_2$ and NaCrS$_2$\cite{Julien2003,Brueesch1971}, considering the higher mass of Te compared to Oxygen and Sulfur.

\begin{figure}[h]
\centering
\includegraphics[width=\linewidth]{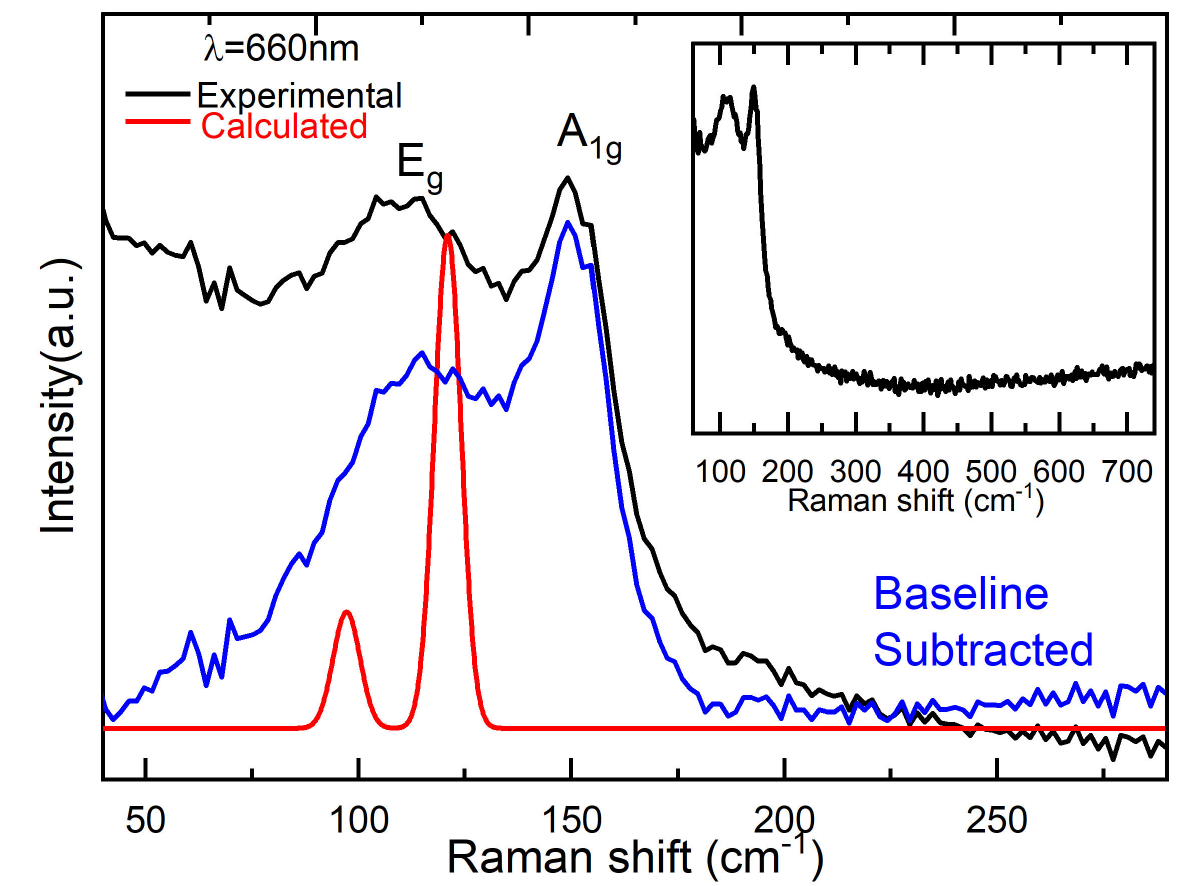}
\caption{Experimental Raman spectrum of AgSbTe$_2$ (black) in comparison with the calculated one (red). The blue spectrum is after a baseline subtraction from the experimental spectrum. The Raman modes symmetries assignment are based on the theoretical calculations. The inset shows an extended Raman shift range of the experimental spectrum.}
\end{figure}

The ability to probe short- and medium-range local atomic arrangements makes nuclear magnetic resonance (NMR) spectroscopy a promising tool for elucidating the structural candidates of AgSbTe$_2$. In particular, the quadrupole moment of nuclei with spin quantum number $I > \frac{1}{2}$ interacts with the surrounding charge distribution via the local electric field gradient (EFG)\cite{Morris2009}. This quadrupolar interaction provides valuable information about the local symmetry and can be used to differentiate between competing structural models\cite{Man2006}. To this end, EFG tensors in their principal axis system were calculated for both the rocksalt-like $Fm\overline{3}m$ and the rhombohedral $R\overline{3}m$ structural candidates of AgSbTe$_2$. 

\begin{figure}[h]
\includegraphics[width=\linewidth]{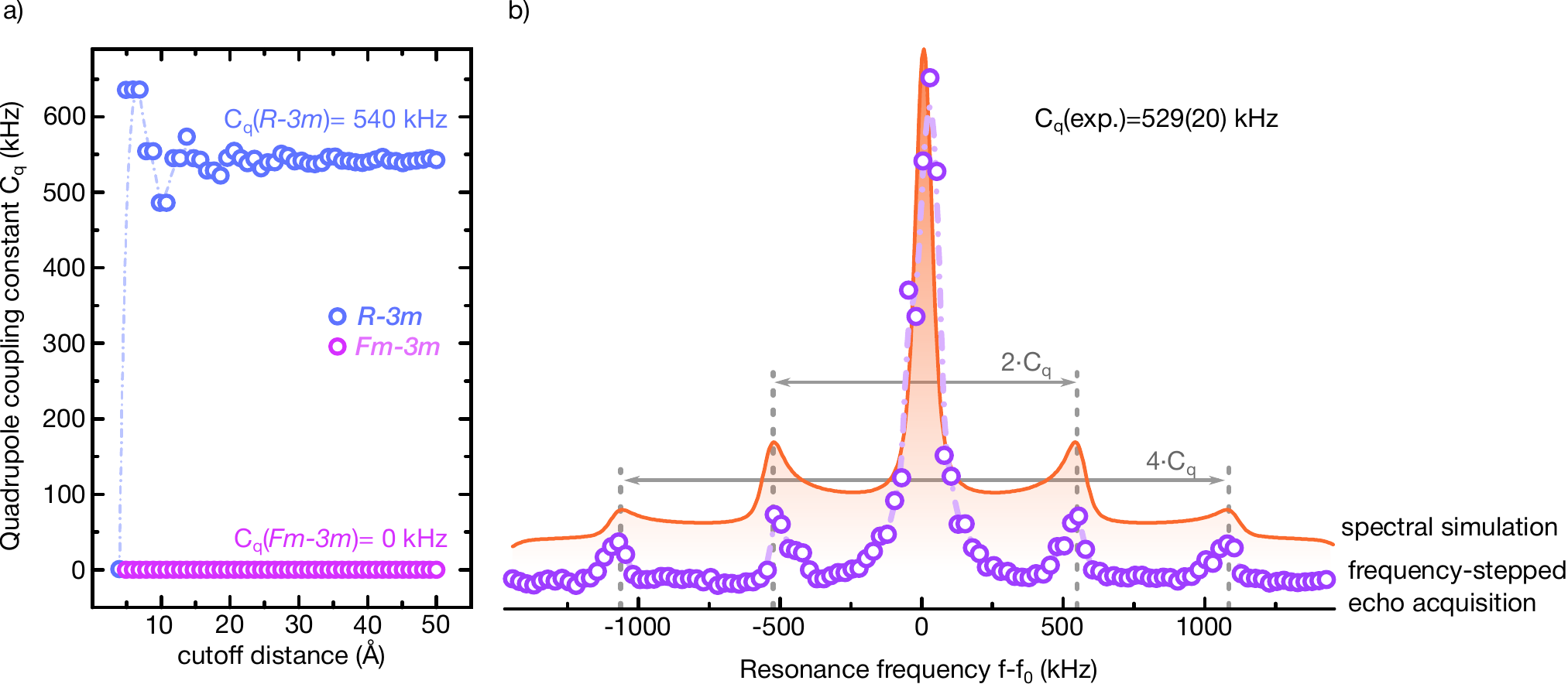}
\caption{(a) Numerical quadrupole coupling constants of $^{121}Sb$ in $Fm-3m$ and $R-3m$ structures as a function of cutoff distance (\AA). Values of 0 and 540 kHz were computed for $Fm-3m$ and $R-3m$, respectively. Blue dashed line is a guide to the eye. b)  Frequency-stepped Hahn-echo acquisition recovered the $I=5/2$ quadrupolar powder pattern (purple). The satellite Pake doublets indicate $C_q(exp.)=529(20)~kHz$. The orange line is the corresponding spectral simulation. Spectral center $f_0$ was 59.83 MHz  at 5.87 T.}
\end{figure}

Figure 4(a) a shows the calculated values of the quadrupolar coupling constant $C_q$ (in kHz) as a function of the cutoff distance from the antimony atoms for both the $Fm\overline{3}m$ and $R\overline{3}m$ structural models. As shown, $C_q$ remains zero for the rocksalt-like structure, indicating complete quenching of the quadrupolar interaction, whereas the rhombohedral structure yields coupling constants around 540~kHz, consistent with an NMR spectrum governed by first-order quadrupolar interaction\cite{Freude1993}. The experimental $^{121}$Sb-NMR spectrum measured at a magnetic field strength of 5.87~T, i shown in Fig. 4(b) . The positions of the spectral singularities associated with the multiple-quantum satellite transitions, i.e., $\ket{\pm 5/2} \leftrightarrow \ket{\pm 3/2}$ and $\ket{\pm 3/2} \leftrightarrow \ket{\pm 1/2}$, indicate a quadrupolar coupling constant of $C_q^\text{(exp.)} = 529(20)$~kHz, in good agreement with the computed quadrupole coupling constant of the rhombohedral $R\overline{3}m$ structural candidate.


The XRD results indicate that all of the previously suggested structures can fit the experimental pattern adequately. The  Rietveld refinements performed for the  various structures  also showed that there are negligible differences between structures  \cite{quarez2005,ko2014}. However, to our knowledge, no ND  results have been reported for AgSbTe$_2$. Only one study reported ND measurements for AgSbSe$_2$, from which it was also not possible to determine cation ordering \cite{Wojciechowski2010}. Our ND Rietveld refinements showed that the L1$_1$ structure exhibits a slightly better fit, for all refinement criteria used in this study; see Table I. However, although these results point toward the L1$_1$ structure,  the relative marginal differences preclude a definitive crystal structure determination.    The relatively large GOF, R-$\chi^2$ and  R$_w$ for all refinements were due to the $\beta$-Ag$_2$Te second phase. If this phase is considered in the refinements, as shown in Table 1 for the case   of the $Fm-3m$-AgSbTe$_2$ + $\beta$-Ag$_2$Te mixture, then these values would be significantly lower, without changing the relative order of the fit quality. 

According to the  ND results, there is an $\approx$4\% volume fraction of $\beta$-Ag$_2$Te as a second phase in the sample. In order to rule out the possibility that  the observed   Raman modes originate from this phase, albeit its  very low amount, the experimental Raman spectrum was also compared to  the  Raman spectra of $\beta$-Ag$_2$Te from previous studies.  The Raman spectrum of bulk polycrystalline $\beta$-Ag$_2$Te consists of two low intensity broad  bands at 111 and 134 cm$^{-1}$ and a broader feature at about 80 cm$^{-1}$\cite{milenov2014}. Also, the Raman spectrum of nano $\beta$-Ag$_2$Te thin film exhibits  a sharp peak at 152 cm$^{-1}$ and a broad peak in the  220 cm$^{-1}$ - 315 cm$^{-1}$ frequency range \cite{pandiaraman2011}. Another study found that the Raman spectrum of 3 nm thick  $\beta$-Ag$_2$Te  consists of Raman bands at   94.3, 121.8, and 141.9 cm$^{-1}$ \cite{sun2023}. 

In our study, only two bands were observed at frequencies clearly distinct from all previously reported ones for various forms of $\beta$-Ag$_2$Te. Moreover, according to previous Raman studies on AgSbTe$_2$, it was established that a high-intensity of the laser excitation  can result in the decomposition of AgSbTe$_2$  and the formation of TeO$_2$, with characteristic strong peaks at 360-400 cm$^{-1}$ and 660 cm$^{-1}$ \cite{milenov2014}. Such peaks are entirely absent in our experimental spectrum; see inset of Fig. 3. From the above discussion, together with the very low quantity of the $\beta$-Ag$_2$Te impurity, it can be accurately concluded that the observed Raman modes originate from the pure AgSbTe$_2$.

The study of the local charge symmetry surrounding the antimony atoms, probed by the presented NMR experiments, demonstrated that a structurally isotropic rocksalt-like $Fm\overline{3}m$ structure would result in a complete quenching of quadrupolar interaction and the observation of a single sharp resonance line, in clear contradiction to experimental observation. Indeed, recorded $^{121}Sb$-NMR spectra show that the antimony spin system in AgSbTe$_2$ is subject to first order quadrupole interaction, strongly favoring the rhombohedral $R\overline3m$ structure, in good accordance with electric field gradient calculations.


Considering the combination of four experimental methods; X-ray and neutron diffraction, Raman and NMR spectroscopy  together  with \textit{ab-initio} DFT calculations, we were able to collect conclusive evidence indicating the rhombohedral $R\overline{3}m$ crystal structure as the most likely structural candidate of the ternary thermoelectric chalcogenide AgSbTe$_2$. Although neutron diffraction refinements are compatible with the proposed model, conclusive structural identification arises from the excellent agreement between experimental and calculated Raman spectra. This assignment is further corroborated by NMR spectroscopy, which independently favors the rhombohedral $R\overline{3}m$ structure. Our study resolves a 60-years old debate about the crystal symmetry of AgSbT$e_2$ and will further facilitate relevant studies exploring and tuning its thermoelectric properties.

The financial support from Guangdong Technion-Israel Institute of Technology (Grant No. ST2100002), MATEC (Grant No. 2022B1212010007, Guangdong Department of Science and Technology), and the Natural Sciences and Engineering Research Council of Canada (NSERC) is acknowledged.  Y.A. would like to acknowledge generous support from the Pazy Research Foundation, Grant No. 2032063.  The authors acknowledge the beamtime at the GPPD granted by the China Spallation Neutron Source (Proposal ID:P1823122900035). This work was also supported by the National Key Research and Development Program of China (2022YFA1402301) and the National Science Foundation of China (42150101). T. Meier acknowledges financial support from Shanghai Key Laboratory Novel Extreme Condition Materials, China (no. 22dz2260800), Shanghai Science and Technology Committee, China (no. 22JC1410300).

The authors have no conflicts to disclose.

\section*{DATA AVAILABILITY } 
The data that support the findings of this study are available from the corresponding author upon reasonable request.

\bibliography{AgSbTe2amb}

\begin{thebibliography}{45}%
\makeatletter
\providecommand \@ifxundefined [1]{%
 \@ifx{#1\undefined}
}%
\providecommand \@ifnum [1]{%
 \ifnum #1\expandafter \@firstoftwo
 \else \expandafter \@secondoftwo
 \fi
}%
\providecommand \@ifx [1]{%
 \ifx #1\expandafter \@firstoftwo
 \else \expandafter \@secondoftwo
 \fi
}%
\providecommand \natexlab [1]{#1}%
\providecommand \enquote  [1]{``#1''}%
\providecommand \bibnamefont  [1]{#1}%
\providecommand \bibfnamefont [1]{#1}%
\providecommand \citenamefont [1]{#1}%
\providecommand \href@noop [0]{\@secondoftwo}%
\providecommand \href [0]{\begingroup \@sanitize@url \@href}%
\providecommand \@href[1]{\@@startlink{#1}\@@href}%
\providecommand \@@href[1]{\endgroup#1\@@endlink}%
\providecommand \@sanitize@url [0]{\catcode `\\12\catcode `\$12\catcode
  `\&12\catcode `\#12\catcode `\^12\catcode `\_12\catcode `\%12\relax}%
\providecommand \@@startlink[1]{}%
\providecommand \@@endlink[0]{}%
\providecommand \url  [0]{\begingroup\@sanitize@url \@url }%
\providecommand \@url [1]{\endgroup\@href {#1}{\urlprefix }}%
\providecommand \urlprefix  [0]{URL }%
\providecommand \Eprint [0]{\href }%
\providecommand \doibase [0]{https://doi.org/}%
\providecommand \selectlanguage [0]{\@gobble}%
\providecommand \bibinfo  [0]{\@secondoftwo}%
\providecommand \bibfield  [0]{\@secondoftwo}%
\providecommand \translation [1]{[#1]}%
\providecommand \BibitemOpen [0]{}%
\providecommand \bibitemStop [0]{}%
\providecommand \bibitemNoStop [0]{.\EOS\space}%
\providecommand \EOS [0]{\spacefactor3000\relax}%
\providecommand \BibitemShut  [1]{\csname bibitem#1\endcsname}%
\let\auto@bib@innerbib\@empty
\bibitem [{\citenamefont {Shi}\ \emph {et~al.}(2020)\citenamefont {Shi},
  \citenamefont {Zou},\ and\ \citenamefont {Chen}}]{shi2020advanced}%
  \BibitemOpen
  \bibfield  {author} {\bibinfo {author} {\bibfnamefont {X.-L.}\ \bibnamefont
  {Shi}}, \bibinfo {author} {\bibfnamefont {J.}~\bibnamefont {Zou}},\ and\
  \bibinfo {author} {\bibfnamefont {Z.-G.}\ \bibnamefont {Chen}},\ }\bibfield
  {title} {\bibinfo {title} {Advanced thermoelectric design: from materials and
  structures to devices},\ }\href@noop {} {\bibfield  {journal} {\bibinfo
  {journal} {Chem. Rev.}\ }\textbf {\bibinfo {volume} {120}},\ \bibinfo {pages}
  {7399} (\bibinfo {year} {2020})}\BibitemShut {NoStop}%
\bibitem [{\citenamefont {Morelli}\ \emph {et~al.}(2008)\citenamefont
  {Morelli}, \citenamefont {Jovovic},\ and\ \citenamefont
  {Heremans}}]{morelli2008}%
  \BibitemOpen
  \bibfield  {author} {\bibinfo {author} {\bibfnamefont {D.}~\bibnamefont
  {Morelli}}, \bibinfo {author} {\bibfnamefont {V.}~\bibnamefont {Jovovic}},\
  and\ \bibinfo {author} {\bibfnamefont {J.}~\bibnamefont {Heremans}},\
  }\bibfield  {title} {\bibinfo {title} {Intrinsically minimal thermal
  conductivity in cubic {I-V-VI}$_2$ semiconductors},\ }\href@noop {}
  {\bibfield  {journal} {\bibinfo  {journal} {Phys. Rev. Lett.}\ }\textbf
  {\bibinfo {volume} {101}},\ \bibinfo {pages} {035901} (\bibinfo {year}
  {2008})}\BibitemShut {NoStop}%
\bibitem [{\citenamefont {Carlton}\ \emph {et~al.}(2014)\citenamefont
  {Carlton}, \citenamefont {De~Armas}, \citenamefont {Ma}, \citenamefont {May},
  \citenamefont {Delaire},\ and\ \citenamefont {Shao-Horn}}]{carlton2014}%
  \BibitemOpen
  \bibfield  {author} {\bibinfo {author} {\bibfnamefont {C.~E.}\ \bibnamefont
  {Carlton}}, \bibinfo {author} {\bibfnamefont {R.}~\bibnamefont {De~Armas}},
  \bibinfo {author} {\bibfnamefont {J.}~\bibnamefont {Ma}}, \bibinfo {author}
  {\bibfnamefont {A.~F.}\ \bibnamefont {May}}, \bibinfo {author} {\bibfnamefont
  {O.}~\bibnamefont {Delaire}},\ and\ \bibinfo {author} {\bibfnamefont
  {Y.}~\bibnamefont {Shao-Horn}},\ }\bibfield  {title} {\bibinfo {title}
  {Natural nanostructure and superlattice nanodomains in {AgSbTe}$_2$},\
  }\href@noop {} {\bibfield  {journal} {\bibinfo  {journal} {J. Appl. Phys.}\
  }\textbf {\bibinfo {volume} {115}} (\bibinfo {year} {2014})}\BibitemShut
  {NoStop}%
\bibitem [{\citenamefont {Hoang}\ \emph {et~al.}(2007)\citenamefont {Hoang},
  \citenamefont {Mahanti}, \citenamefont {Salvador},\ and\ \citenamefont
  {Kanatzidis}}]{hoang2007}%
  \BibitemOpen
  \bibfield  {author} {\bibinfo {author} {\bibfnamefont {K.}~\bibnamefont
  {Hoang}}, \bibinfo {author} {\bibfnamefont {S.~D.}\ \bibnamefont {Mahanti}},
  \bibinfo {author} {\bibfnamefont {J.~R.}\ \bibnamefont {Salvador}},\ and\
  \bibinfo {author} {\bibfnamefont {M.~G.}\ \bibnamefont {Kanatzidis}},\
  }\bibfield  {title} {\bibinfo {title} {Atomic ordering and gap formation in
  {Ag-Sb-Based} ternary chalcogenides},\ }\href
  {https://doi.org/10.1103/PhysRevLett.99.156403} {\bibfield  {journal}
  {\bibinfo  {journal} {Phys. Rev. Lett.}\ }\textbf {\bibinfo {volume} {99}},\
  \bibinfo {pages} {156403} (\bibinfo {year} {2007})}\BibitemShut {NoStop}%
\bibitem [{\citenamefont {Jovovic}\ and\ \citenamefont
  {Heremans}(2008)}]{jovovic2008}%
  \BibitemOpen
  \bibfield  {author} {\bibinfo {author} {\bibfnamefont {V.}~\bibnamefont
  {Jovovic}}\ and\ \bibinfo {author} {\bibfnamefont {J.~P.}\ \bibnamefont
  {Heremans}},\ }\bibfield  {title} {\bibinfo {title} {Measurements of the
  energy band gap and valence band structure of {AgSbTe}$_2$},\ }\href
  {https://doi.org/10.1103/PhysRevB.77.245204} {\bibfield  {journal} {\bibinfo
  {journal} {Phys. Rev. B}\ }\textbf {\bibinfo {volume} {77}},\ \bibinfo
  {pages} {245204} (\bibinfo {year} {2008})}\BibitemShut {NoStop}%
\bibitem [{\citenamefont {Roychowdhury}\ \emph {et~al.}(2021)\citenamefont
  {Roychowdhury}, \citenamefont {Ghosh}, \citenamefont {Arora}, \citenamefont
  {Samanta}, \citenamefont {Xie}, \citenamefont {Singh}, \citenamefont {Soni},
  \citenamefont {He}, \citenamefont {Waghmare},\ and\ \citenamefont
  {Biswas}}]{Roychowdhury2021}%
  \BibitemOpen
  \bibfield  {author} {\bibinfo {author} {\bibfnamefont {S.}~\bibnamefont
  {Roychowdhury}}, \bibinfo {author} {\bibfnamefont {T.}~\bibnamefont {Ghosh}},
  \bibinfo {author} {\bibfnamefont {R.}~\bibnamefont {Arora}}, \bibinfo
  {author} {\bibfnamefont {M.}~\bibnamefont {Samanta}}, \bibinfo {author}
  {\bibfnamefont {L.}~\bibnamefont {Xie}}, \bibinfo {author} {\bibfnamefont
  {N.~K.}\ \bibnamefont {Singh}}, \bibinfo {author} {\bibfnamefont
  {A.}~\bibnamefont {Soni}}, \bibinfo {author} {\bibfnamefont {J.}~\bibnamefont
  {He}}, \bibinfo {author} {\bibfnamefont {U.~V.}\ \bibnamefont {Waghmare}},\
  and\ \bibinfo {author} {\bibfnamefont {K.}~\bibnamefont {Biswas}},\
  }\bibfield  {title} {\bibinfo {title} {Enhanced atomic ordering leads to high
  thermoelectric performance in {AgSbTe}$_2$},\ }\href
  {https://doi.org/10.1126/science.abb3517} {\bibfield  {journal} {\bibinfo
  {journal} {Science}\ }\textbf {\bibinfo {volume} {371}},\ \bibinfo {pages}
  {722} (\bibinfo {year} {2021})}\BibitemShut {NoStop}%
\bibitem [{\citenamefont {Geller}\ and\ \citenamefont
  {Wernick}(1959)}]{geller1959}%
  \BibitemOpen
  \bibfield  {author} {\bibinfo {author} {\bibfnamefont {S.}~\bibnamefont
  {Geller}}\ and\ \bibinfo {author} {\bibfnamefont {J.~H.}\ \bibnamefont
  {Wernick}},\ }\bibfield  {title} {\bibinfo {title} {Ternary semiconducting
  compounds with sodium chloride-like structure: {AgSbSe}$_2$, {AgSbTe}$_2$,
  {AgBiS}$_2$, {AgBiSe}$_2$},\ }\href
  {https://doi.org/10.1107/S0365110X59000135} {\bibfield  {journal} {\bibinfo
  {journal} {Acta Cryst. Sect. A}\ }\textbf {\bibinfo {volume} {12}},\ \bibinfo
  {pages} {46} (\bibinfo {year} {1959})}\BibitemShut {NoStop}%
\bibitem [{\citenamefont {Quarez}\ \emph {et~al.}(2005)\citenamefont {Quarez},
  \citenamefont {Hsu}, \citenamefont {Pcionek}, \citenamefont {Frangis},
  \citenamefont {Polychroniadis},\ and\ \citenamefont
  {Kanatzidis}}]{quarez2005}%
  \BibitemOpen
  \bibfield  {author} {\bibinfo {author} {\bibfnamefont {E.}~\bibnamefont
  {Quarez}}, \bibinfo {author} {\bibfnamefont {K.-F.}\ \bibnamefont {Hsu}},
  \bibinfo {author} {\bibfnamefont {R.}~\bibnamefont {Pcionek}}, \bibinfo
  {author} {\bibfnamefont {N.}~\bibnamefont {Frangis}}, \bibinfo {author}
  {\bibfnamefont {E.~K.}\ \bibnamefont {Polychroniadis}},\ and\ \bibinfo
  {author} {\bibfnamefont {M.~G.}\ \bibnamefont {Kanatzidis}},\ }\bibfield
  {title} {\bibinfo {title} {Nanostructuring, compositional fluctuations, and
  atomic ordering in the thermoelectric materials {AgPb}$_m${Sb}{Te}$_{2+m}$.
  the myth of solid solutions},\ }\href {https://doi.org/10.1021/ja051653o}
  {\bibfield  {journal} {\bibinfo  {journal} {J. Am. Chem. Soc.}\ }\textbf
  {\bibinfo {volume} {127}},\ \bibinfo {pages} {9177} (\bibinfo {year}
  {2005})}\BibitemShut {NoStop}%
\bibitem [{\citenamefont {Barabash}\ \emph {et~al.}(2008)\citenamefont
  {Barabash}, \citenamefont {Ozolins},\ and\ \citenamefont
  {Wolverton}}]{barabash2008}%
  \BibitemOpen
  \bibfield  {author} {\bibinfo {author} {\bibfnamefont {S.}~\bibnamefont
  {Barabash}}, \bibinfo {author} {\bibfnamefont {V.}~\bibnamefont {Ozolins}},\
  and\ \bibinfo {author} {\bibfnamefont {C.}~\bibnamefont {Wolverton}},\
  }\bibfield  {title} {\bibinfo {title} {First-principles theory of competing
  order types, phase separation, and phonon spectra in thermoelectric
  {AgPb}$_m$ {SbTe}$_{m+2}$ alloys},\ }\href@noop {} {\bibfield  {journal}
  {\bibinfo  {journal} {Phys. Rev. Lett.}\ }\textbf {\bibinfo {volume} {101}},\
  \bibinfo {pages} {155704} (\bibinfo {year} {2008})}\BibitemShut {NoStop}%
\bibitem [{\citenamefont {Rezaei}\ \emph {et~al.}(2014)\citenamefont {Rezaei},
  \citenamefont {Hashemifar},\ and\ \citenamefont {Akbarzadeh}}]{Rezaei2014}%
  \BibitemOpen
  \bibfield  {author} {\bibinfo {author} {\bibfnamefont {N.}~\bibnamefont
  {Rezaei}}, \bibinfo {author} {\bibfnamefont {S.~J.}\ \bibnamefont
  {Hashemifar}},\ and\ \bibinfo {author} {\bibfnamefont {H.}~\bibnamefont
  {Akbarzadeh}},\ }\bibfield  {title} {\bibinfo {title} {Thermoelectric
  properties of {AgSbTe}$_2$ from first-principles calculations},\ }\href
  {https://doi.org/10.1063/1.4895062} {\bibfield  {journal} {\bibinfo
  {journal} {J. Appl. Phys.}\ }\textbf {\bibinfo {volume} {116}},\ \bibinfo
  {pages} {103705} (\bibinfo {year} {2014})}\BibitemShut {NoStop}%
\bibitem [{\citenamefont {Szczypka}\ and\ \citenamefont
  {Koleżyński}(2018)}]{Szczypka2018}%
  \BibitemOpen
  \bibfield  {author} {\bibinfo {author} {\bibfnamefont {W.}~\bibnamefont
  {Szczypka}}\ and\ \bibinfo {author} {\bibfnamefont {A.}~\bibnamefont
  {Koleżyński}},\ }\bibfield  {title} {\bibinfo {title} {Theoretical studies
  of cation sublattice ordering in {AgSbTe}$_2$ and {AgSbSe}$_2$ – electron
  density topology and bonding properties},\ }\href
  {https://doi.org/https://doi.org/10.1016/j.jallcom.2017.10.199} {\bibfield
  {journal} {\bibinfo  {journal} {J. Alloys Compd.}\ }\textbf {\bibinfo
  {volume} {732}},\ \bibinfo {pages} {293} (\bibinfo {year}
  {2018})}\BibitemShut {NoStop}%
\bibitem [{\citenamefont {Amouyal}(2013)}]{Amouyal2013}%
  \BibitemOpen
  \bibfield  {author} {\bibinfo {author} {\bibfnamefont {Y.}~\bibnamefont
  {Amouyal}},\ }\bibfield  {title} {\bibinfo {title} {On the role of lanthanum
  substitution defects in reducing lattice thermal conductivity of the
  {AgSbTe}$_2$ ({P4/mmm}) thermoelectric compound for energy conversion
  applications},\ }\href
  {https://doi.org/https://doi.org/10.1016/j.commatsci.2013.05.027} {\bibfield
  {journal} {\bibinfo  {journal} {Comput. Mater. Sci.}\ }\textbf {\bibinfo
  {volume} {78}},\ \bibinfo {pages} {98} (\bibinfo {year} {2013})}\BibitemShut
  {NoStop}%
\bibitem [{\citenamefont {Amouyal}(2014)}]{Amouyal2014}%
  \BibitemOpen
  \bibfield  {author} {\bibinfo {author} {\bibfnamefont {Y.}~\bibnamefont
  {Amouyal}},\ }\bibfield  {title} {\bibinfo {title} {Reducing lattice thermal
  conductivity of the thermoelectric compound {AgSbTe}$_2$ {(P4/mmm}) by
  lanthanum substitution: Computational and experimental approaches},\ }\href
  {https://doi.org/10.1007/s11664-014-3145-y} {\bibfield  {journal} {\bibinfo
  {journal} {J. Electron. Mater.}\ }\textbf {\bibinfo {volume} {43}},\ \bibinfo
  {pages} {3772} (\bibinfo {year} {2014})}\BibitemShut {NoStop}%
\bibitem [{\citenamefont {Amouyal}(2016)}]{Amouyal2016}%
  \BibitemOpen
  \bibfield  {author} {\bibinfo {author} {\bibfnamefont {Y.}~\bibnamefont
  {Amouyal}},\ }\bibfield  {title} {\bibinfo {title}
  {Silver-antimony-telluride: From first-principles calculations to
  thermoelectric applications},\ }in\ \href {https://doi.org/10.5772/66086}
  {\emph {\bibinfo {booktitle} {Thermoelectrics for Power Generation}}},\
  \bibinfo {editor} {edited by\ \bibinfo {editor} {\bibfnamefont
  {S.}~\bibnamefont {Skipidarov}}\ and\ \bibinfo {editor} {\bibfnamefont
  {M.}~\bibnamefont {Nikitin}}}\ (\bibinfo  {publisher} {IntechOpen},\ \bibinfo
  {address} {Rijeka},\ \bibinfo {year} {2016})\ Chap.~\bibinfo {chapter}
  {7}\BibitemShut {NoStop}%
\bibitem [{\citenamefont {He}\ \emph {et~al.}(2023)\citenamefont {He},
  \citenamefont {Deng}, \citenamefont {Shen}, \citenamefont {Chen},
  \citenamefont {Lu}, \citenamefont {Tan}, \citenamefont {Zheng}, \citenamefont
  {Hao}, \citenamefont {Zhao},\ and\ \citenamefont {Ma}}]{He2023}%
  \BibitemOpen
  \bibfield  {author} {\bibinfo {author} {\bibfnamefont {L.}~\bibnamefont
  {He}}, \bibinfo {author} {\bibfnamefont {S.}~\bibnamefont {Deng}}, \bibinfo
  {author} {\bibfnamefont {F.}~\bibnamefont {Shen}}, \bibinfo {author}
  {\bibfnamefont {J.}~\bibnamefont {Chen}}, \bibinfo {author} {\bibfnamefont
  {H.}~\bibnamefont {Lu}}, \bibinfo {author} {\bibfnamefont {Z.}~\bibnamefont
  {Tan}}, \bibinfo {author} {\bibfnamefont {H.}~\bibnamefont {Zheng}}, \bibinfo
  {author} {\bibfnamefont {J.}~\bibnamefont {Hao}}, \bibinfo {author}
  {\bibfnamefont {D.}~\bibnamefont {Zhao}},\ and\ \bibinfo {author}
  {\bibfnamefont {Q.}~\bibnamefont {Ma}},\ }\bibfield  {title} {\bibinfo
  {title} {The performance of general purpose powder diffractometer at
  {CSNS}},\ }\href {https://doi.org/https://doi.org/10.1016/j.nima.2023.168414}
  {\bibfield  {journal} {\bibinfo  {journal} {Nucl. Instrum. Methods Phys. Res.
  Sect. A}\ }\textbf {\bibinfo {volume} {1054}},\ \bibinfo {pages} {168414}
  (\bibinfo {year} {2023})}\BibitemShut {NoStop}%
\bibitem [{\citenamefont {Hao}\ \emph {et~al.}(2023)\citenamefont {Hao},
  \citenamefont {Tan}, \citenamefont {Lu}, \citenamefont {Deng}, \citenamefont
  {Shen}, \citenamefont {Zhao}, \citenamefont {Zheng}, \citenamefont {Ma},
  \citenamefont {Chen},\ and\ \citenamefont {He}}]{Hao2023}%
  \BibitemOpen
  \bibfield  {author} {\bibinfo {author} {\bibfnamefont {J.}~\bibnamefont
  {Hao}}, \bibinfo {author} {\bibfnamefont {Z.}~\bibnamefont {Tan}}, \bibinfo
  {author} {\bibfnamefont {H.}~\bibnamefont {Lu}}, \bibinfo {author}
  {\bibfnamefont {S.}~\bibnamefont {Deng}}, \bibinfo {author} {\bibfnamefont
  {F.}~\bibnamefont {Shen}}, \bibinfo {author} {\bibfnamefont {D.}~\bibnamefont
  {Zhao}}, \bibinfo {author} {\bibfnamefont {H.}~\bibnamefont {Zheng}},
  \bibinfo {author} {\bibfnamefont {Q.}~\bibnamefont {Ma}}, \bibinfo {author}
  {\bibfnamefont {J.}~\bibnamefont {Chen}},\ and\ \bibinfo {author}
  {\bibfnamefont {L.}~\bibnamefont {He}},\ }\bibfield  {title} {\bibinfo
  {title} {Residual stress measurement system of the general purpose powder
  diffractometer at {CSNS}},\ }\href
  {https://doi.org/https://doi.org/10.1016/j.nima.2023.168532} {\bibfield
  {journal} {\bibinfo  {journal} {Nucl. Instrum. Methods Phys. Res. Sect. A}\
  }\textbf {\bibinfo {volume} {1055}},\ \bibinfo {pages} {168532} (\bibinfo
  {year} {2023})}\BibitemShut {NoStop}%
\bibitem [{\citenamefont {Hinton}\ \emph {et~al.}(2019)\citenamefont {Hinton},
  \citenamefont {Clarke}, \citenamefont {Steele}, \citenamefont {Kuo},
  \citenamefont {Greenberg}, \citenamefont {Prakapenka}, \citenamefont {Kunz},
  \citenamefont {Kroonblawd},\ and\ \citenamefont {Stavrou}}]{Hinton2019}%
  \BibitemOpen
  \bibfield  {author} {\bibinfo {author} {\bibfnamefont {J.~K.}\ \bibnamefont
  {Hinton}}, \bibinfo {author} {\bibfnamefont {S.~M.}\ \bibnamefont {Clarke}},
  \bibinfo {author} {\bibfnamefont {B.~A.}\ \bibnamefont {Steele}}, \bibinfo
  {author} {\bibfnamefont {I.-F.~W.}\ \bibnamefont {Kuo}}, \bibinfo {author}
  {\bibfnamefont {E.}~\bibnamefont {Greenberg}}, \bibinfo {author}
  {\bibfnamefont {V.~B.}\ \bibnamefont {Prakapenka}}, \bibinfo {author}
  {\bibfnamefont {M.}~\bibnamefont {Kunz}}, \bibinfo {author} {\bibfnamefont
  {M.~P.}\ \bibnamefont {Kroonblawd}},\ and\ \bibinfo {author} {\bibfnamefont
  {E.}~\bibnamefont {Stavrou}},\ }\bibfield  {title} {\bibinfo {title} {Effects
  of pressure on the structure and lattice dynamics of $\alpha$-glycine: a
  combined experimental and theoretical study},\ }\href
  {https://doi.org/10.1039/C8CE02123F} {\bibfield  {journal} {\bibinfo
  {journal} {CrystEngComm}\ }\textbf {\bibinfo {volume} {21}},\ \bibinfo
  {pages} {4457} (\bibinfo {year} {2019})}\BibitemShut {NoStop}%
\bibitem [{\citenamefont {Hahn}(1950)}]{Hahn1950}%
  \BibitemOpen
  \bibfield  {author} {\bibinfo {author} {\bibfnamefont {E.~L.}\ \bibnamefont
  {Hahn}},\ }\bibfield  {title} {\bibinfo {title} {Spin echoes},\ }\href
  {https://doi.org/10.1103/PhysRev.80.580} {\bibfield  {journal} {\bibinfo
  {journal} {Phys. Rev.}\ }\textbf {\bibinfo {volume} {80}},\ \bibinfo {pages}
  {580} (\bibinfo {year} {1950})}\BibitemShut {NoStop}%
\bibitem [{\citenamefont {Man}(2006)}]{Man2006}%
  \BibitemOpen
  \bibfield  {author} {\bibinfo {author} {\bibfnamefont {P.~P.}\ \bibnamefont
  {Man}},\ }\bibinfo {title} {Quadrupole couplings in nuclear magnetic
  resonance, general}\ (\bibinfo  {publisher} {John Wiley \& Sons, Ltd},\
  \bibinfo {year} {2006})\BibitemShut {NoStop}%
\bibitem [{\citenamefont {Haase}(1995)}]{Haase1995}%
  \BibitemOpen
  \bibfield  {author} {\bibinfo {author} {\bibfnamefont {J.}~\bibnamefont
  {Haase}},\ }\emph {\bibinfo {title} {{Magnetische Kernresonanz bei
  Quadrupolwechselwirkung in Festkoerpern}}},\ \href@noop {} {Ph.D. thesis},\
  \bibinfo  {school} {Leipzig University} (\bibinfo {year} {1995})\BibitemShut
  {NoStop}%
\bibitem [{\citenamefont {Hughes}\ and\ \citenamefont
  {Harris}(2016)}]{Hughes2016}%
  \BibitemOpen
  \bibfield  {author} {\bibinfo {author} {\bibfnamefont {C.~E.}\ \bibnamefont
  {Hughes}}\ and\ \bibinfo {author} {\bibfnamefont {K.~D.~M.}\ \bibnamefont
  {Harris}},\ }\bibfield  {title} {\bibinfo {title} {Calculation of solid-state
  nmr lineshapes using contour analysis.},\ }\href@noop {} {\bibfield
  {journal} {\bibinfo  {journal} {olid State Nucl. Magn. Reson.}\ }\textbf
  {\bibinfo {volume} {80}},\ \bibinfo {pages} {7} (\bibinfo {year}
  {2016})}\BibitemShut {NoStop}%
\bibitem [{\citenamefont {Baroni}\ \emph {et~al.}(2001)\citenamefont {Baroni},
  \citenamefont {De~Gironcoli}, \citenamefont {Dal~Corso},\ and\ \citenamefont
  {Giannozzi}}]{baroni2001}%
  \BibitemOpen
  \bibfield  {author} {\bibinfo {author} {\bibfnamefont {S.}~\bibnamefont
  {Baroni}}, \bibinfo {author} {\bibfnamefont {S.}~\bibnamefont
  {De~Gironcoli}}, \bibinfo {author} {\bibfnamefont {A.}~\bibnamefont
  {Dal~Corso}},\ and\ \bibinfo {author} {\bibfnamefont {P.}~\bibnamefont
  {Giannozzi}},\ }\bibfield  {title} {\bibinfo {title} {Phonons and related
  crystal properties from density-functional perturbation theory},\ }\href@noop
  {} {\bibfield  {journal} {\bibinfo  {journal} {Rev. Modern Phys.}\ }\textbf
  {\bibinfo {volume} {73}},\ \bibinfo {pages} {515} (\bibinfo {year}
  {2001})}\BibitemShut {NoStop}%
\bibitem [{\citenamefont {Perdew}\ and\ \citenamefont
  {Wang}(1992)}]{perdew1992}%
  \BibitemOpen
  \bibfield  {author} {\bibinfo {author} {\bibfnamefont {J.~P.}\ \bibnamefont
  {Perdew}}\ and\ \bibinfo {author} {\bibfnamefont {Y.}~\bibnamefont {Wang}},\
  }\bibfield  {title} {\bibinfo {title} {Accurate and simple analytic
  representation of the electron-gas correlation energy},\ }\href@noop {}
  {\bibfield  {journal} {\bibinfo  {journal} {Phys. Rev. B}\ }\textbf {\bibinfo
  {volume} {45}},\ \bibinfo {pages} {13244} (\bibinfo {year}
  {1992})}\BibitemShut {NoStop}%
\bibitem [{\citenamefont {Troullier}\ and\ \citenamefont
  {Martins}(1991)}]{troullier1991}%
  \BibitemOpen
  \bibfield  {author} {\bibinfo {author} {\bibfnamefont {N.}~\bibnamefont
  {Troullier}}\ and\ \bibinfo {author} {\bibfnamefont {J.~L.}\ \bibnamefont
  {Martins}},\ }\bibfield  {title} {\bibinfo {title} {Efficient
  pseudopotentials for plane-wave calculations},\ }\href@noop {} {\bibfield
  {journal} {\bibinfo  {journal} {Phys. Rev. B}\ }\textbf {\bibinfo {volume}
  {43}},\ \bibinfo {pages} {1993} (\bibinfo {year} {1991})}\BibitemShut
  {NoStop}%
\bibitem [{\citenamefont {Monkhorst}\ and\ \citenamefont
  {Pack}(1976)}]{monkhorst1976}%
  \BibitemOpen
  \bibfield  {author} {\bibinfo {author} {\bibfnamefont {H.~J.}\ \bibnamefont
  {Monkhorst}}\ and\ \bibinfo {author} {\bibfnamefont {J.~D.}\ \bibnamefont
  {Pack}},\ }\bibfield  {title} {\bibinfo {title} {Special points for
  {B}rillouin-zone integrations},\ }\href@noop {} {\bibfield  {journal}
  {\bibinfo  {journal} {Phys. Rev. B}\ }\textbf {\bibinfo {volume} {13}},\
  \bibinfo {pages} {5188} (\bibinfo {year} {1976})}\BibitemShut {NoStop}%
\bibitem [{\citenamefont {Giannozzi}\ \emph {et~al.}(2009)\citenamefont
  {Giannozzi}, \citenamefont {Baroni}, \citenamefont {Bonini}, \citenamefont
  {Calandra}, \citenamefont {Car}, \citenamefont {Cavazzoni}, \citenamefont
  {Ceresoli}, \citenamefont {Chiarotti}, \citenamefont {Cococcioni},
  \citenamefont {Dabo} \emph {et~al.}}]{giannozzi2009}%
  \BibitemOpen
  \bibfield  {author} {\bibinfo {author} {\bibfnamefont {P.}~\bibnamefont
  {Giannozzi}}, \bibinfo {author} {\bibfnamefont {S.}~\bibnamefont {Baroni}},
  \bibinfo {author} {\bibfnamefont {N.}~\bibnamefont {Bonini}}, \bibinfo
  {author} {\bibfnamefont {M.}~\bibnamefont {Calandra}}, \bibinfo {author}
  {\bibfnamefont {R.}~\bibnamefont {Car}}, \bibinfo {author} {\bibfnamefont
  {C.}~\bibnamefont {Cavazzoni}}, \bibinfo {author} {\bibfnamefont
  {D.}~\bibnamefont {Ceresoli}}, \bibinfo {author} {\bibfnamefont {G.~L.}\
  \bibnamefont {Chiarotti}}, \bibinfo {author} {\bibfnamefont {M.}~\bibnamefont
  {Cococcioni}}, \bibinfo {author} {\bibfnamefont {I.}~\bibnamefont {Dabo}},
  \emph {et~al.},\ }\bibfield  {title} {\bibinfo {title} {Quantum espresso: a
  modular and open-source software project for quantumsimulations of
  materials},\ }\href@noop {} {\bibfield  {journal} {\bibinfo  {journal} {J.
  Phys. Condens. Matter.}\ }\textbf {\bibinfo {volume} {21}},\ \bibinfo {pages}
  {395502} (\bibinfo {year} {2009})}\BibitemShut {NoStop}%
\bibitem [{\citenamefont {Giannozzi}\ \emph {et~al.}(2017)\citenamefont
  {Giannozzi}, \citenamefont {Andreussi}, \citenamefont {Brumme}, \citenamefont
  {Bunau}, \citenamefont {Nardelli}, \citenamefont {Calandra}, \citenamefont
  {Car}, \citenamefont {Cavazzoni}, \citenamefont {Ceresoli}, \citenamefont
  {Cococcioni} \emph {et~al.}}]{giannozzi2017}%
  \BibitemOpen
  \bibfield  {author} {\bibinfo {author} {\bibfnamefont {P.}~\bibnamefont
  {Giannozzi}}, \bibinfo {author} {\bibfnamefont {O.}~\bibnamefont
  {Andreussi}}, \bibinfo {author} {\bibfnamefont {T.}~\bibnamefont {Brumme}},
  \bibinfo {author} {\bibfnamefont {O.}~\bibnamefont {Bunau}}, \bibinfo
  {author} {\bibfnamefont {M.~B.}\ \bibnamefont {Nardelli}}, \bibinfo {author}
  {\bibfnamefont {M.}~\bibnamefont {Calandra}}, \bibinfo {author}
  {\bibfnamefont {R.}~\bibnamefont {Car}}, \bibinfo {author} {\bibfnamefont
  {C.}~\bibnamefont {Cavazzoni}}, \bibinfo {author} {\bibfnamefont
  {D.}~\bibnamefont {Ceresoli}}, \bibinfo {author} {\bibfnamefont
  {M.}~\bibnamefont {Cococcioni}}, \emph {et~al.},\ }\bibfield  {title}
  {\bibinfo {title} {Advanced capabilities for materials modelling with quantum
  espresso},\ }\href@noop {} {\bibfield  {journal} {\bibinfo  {journal} {J.
  Phys. Condens. Matter.}\ }\textbf {\bibinfo {volume} {29}},\ \bibinfo {pages}
  {465901} (\bibinfo {year} {2017})}\BibitemShut {NoStop}%
\bibitem [{\citenamefont {Kraus}\ and\ \citenamefont
  {Nolze}(1996)}]{Kraus1996}%
  \BibitemOpen
  \bibfield  {author} {\bibinfo {author} {\bibfnamefont {W.}~\bibnamefont
  {Kraus}}\ and\ \bibinfo {author} {\bibfnamefont {G.}~\bibnamefont {Nolze}},\
  }\bibfield  {title} {\bibinfo {title} {{POWDER} {CELL}–a program for the
  representation and manipulation of crystal structures and calculation of the
  resulting {X}-ray powder patterns},\ }\href@noop {} {\bibfield  {journal}
  {\bibinfo  {journal} {J. Appl. Crystallogr.}\ }\textbf {\bibinfo {volume}
  {29}},\ \bibinfo {pages} {301} (\bibinfo {year} {1996})}\BibitemShut
  {NoStop}%
\bibitem [{\citenamefont {Hu}\ \emph {et~al.}(2025)\citenamefont {Hu},
  \citenamefont {Yuan}, \citenamefont {Li}, \citenamefont {Wang}, \citenamefont
  {Li}, \citenamefont {Jiang}, \citenamefont {Shuai},\ and\ \citenamefont
  {Hou}}]{hu2025}%
  \BibitemOpen
  \bibfield  {author} {\bibinfo {author} {\bibfnamefont {Z.}~\bibnamefont
  {Hu}}, \bibinfo {author} {\bibfnamefont {M.}~\bibnamefont {Yuan}}, \bibinfo
  {author} {\bibfnamefont {W.}~\bibnamefont {Li}}, \bibinfo {author}
  {\bibfnamefont {S.}~\bibnamefont {Wang}}, \bibinfo {author} {\bibfnamefont
  {J.}~\bibnamefont {Li}}, \bibinfo {author} {\bibfnamefont {J.}~\bibnamefont
  {Jiang}}, \bibinfo {author} {\bibfnamefont {J.}~\bibnamefont {Shuai}},\ and\
  \bibinfo {author} {\bibfnamefont {Y.}~\bibnamefont {Hou}},\ }\bibfield
  {title} {\bibinfo {title} {Enhanced thermoelectric performance in pristine
  {AgSbTe}$_2$ compound via rational design of {Ag}$_2${Te} formation},\
  }\href@noop {} {\bibfield  {journal} {\bibinfo  {journal} {Acta Materialia}\
  ,\ \bibinfo {pages} {120985}} (\bibinfo {year} {2025})}\BibitemShut {NoStop}%
\bibitem [{\citenamefont {Kim}\ \emph {et~al.}(2024)\citenamefont {Kim},
  \citenamefont {Yun}, \citenamefont {Cha}, \citenamefont {Byeon},
  \citenamefont {Park}, \citenamefont {Jin}, \citenamefont {Kim}, \citenamefont
  {Kim}, \citenamefont {Park}, \citenamefont {Jang} \emph {et~al.}}]{kim2024}%
  \BibitemOpen
  \bibfield  {author} {\bibinfo {author} {\bibfnamefont {J.~H.}\ \bibnamefont
  {Kim}}, \bibinfo {author} {\bibfnamefont {J.~H.}\ \bibnamefont {Yun}},
  \bibinfo {author} {\bibfnamefont {S.}~\bibnamefont {Cha}}, \bibinfo {author}
  {\bibfnamefont {S.}~\bibnamefont {Byeon}}, \bibinfo {author} {\bibfnamefont
  {J.}~\bibnamefont {Park}}, \bibinfo {author} {\bibfnamefont {H.}~\bibnamefont
  {Jin}}, \bibinfo {author} {\bibfnamefont {S.}~\bibnamefont {Kim}}, \bibinfo
  {author} {\bibfnamefont {S.-J.}\ \bibnamefont {Kim}}, \bibinfo {author}
  {\bibfnamefont {J.}~\bibnamefont {Park}}, \bibinfo {author} {\bibfnamefont
  {J.}~\bibnamefont {Jang}}, \emph {et~al.},\ }\bibfield  {title} {\bibinfo
  {title} {Enhancement of phase stability and thermoelectric performance of
  meta-stable {AgSbTe}$_2$ by thermal cycling process},\ }\href@noop {}
  {\bibfield  {journal} {\bibinfo  {journal} {Adv. Funct. Mater.}\ }\textbf
  {\bibinfo {volume} {34}},\ \bibinfo {pages} {2404886} (\bibinfo {year}
  {2024})}\BibitemShut {NoStop}%
\bibitem [{\citenamefont {Toby}\ and\ \citenamefont
  {Von~Dreele}(2013)}]{Toby2013}%
  \BibitemOpen
  \bibfield  {author} {\bibinfo {author} {\bibfnamefont {B.~H.}\ \bibnamefont
  {Toby}}\ and\ \bibinfo {author} {\bibfnamefont {R.~B.}\ \bibnamefont
  {Von~Dreele}},\ }\bibfield  {title} {\bibinfo {title} {{{\it GSAS-II}: the
  genesis of a modern open-source all purpose crystallography software
  package}},\ }\href {https://doi.org/10.1107/S0021889813003531} {\bibfield
  {journal} {\bibinfo  {journal} {J. Appl. Crystallogr.}\ }\textbf {\bibinfo
  {volume} {46}},\ \bibinfo {pages} {544} (\bibinfo {year} {2013})}\BibitemShut
  {NoStop}%
\bibitem [{sup()}]{supp}%
  \BibitemOpen
  \href@noop {} {}\bibinfo {howpublished} {See Supplemental Material at , which
  includes supplemental figures 1-2}\BibitemShut {NoStop}%
\bibitem [{\citenamefont {Sears}(1992)}]{Sears1992}%
  \BibitemOpen
  \bibfield  {author} {\bibinfo {author} {\bibfnamefont {V.~F.}\ \bibnamefont
  {Sears}},\ }\bibfield  {title} {\bibinfo {title} {Neutron scattering lengths
  and cross sections},\ }\href {https://doi.org/10.1080/10448639208218770}
  {\bibfield  {journal} {\bibinfo  {journal} {Neutron News}\ }\textbf {\bibinfo
  {volume} {3}},\ \bibinfo {pages} {26} (\bibinfo {year} {1992})}\BibitemShut
  {NoStop}%
\bibitem [{\citenamefont {Kroumova}\ \emph {et~al.}(2003)\citenamefont
  {Kroumova}, \citenamefont {Aroyo}, \citenamefont {Perez-Mato}, \citenamefont
  {Kirov}, \citenamefont {Capillas}, \citenamefont {Ivantchev},\ and\
  \citenamefont {and}}]{Kroumova2003}%
  \BibitemOpen
  \bibfield  {author} {\bibinfo {author} {\bibfnamefont {E.}~\bibnamefont
  {Kroumova}}, \bibinfo {author} {\bibfnamefont {M.}~\bibnamefont {Aroyo}},
  \bibinfo {author} {\bibfnamefont {J.}~\bibnamefont {Perez-Mato}}, \bibinfo
  {author} {\bibfnamefont {A.}~\bibnamefont {Kirov}}, \bibinfo {author}
  {\bibfnamefont {C.}~\bibnamefont {Capillas}}, \bibinfo {author}
  {\bibfnamefont {S.}~\bibnamefont {Ivantchev}},\ and\ \bibinfo {author}
  {\bibfnamefont {H.~W.}\ \bibnamefont {and}},\ }\bibfield  {title} {\bibinfo
  {title} {Bilbao crystallographic server : Useful databases and tools for
  phase-transition studies},\ }\href
  {https://doi.org/10.1080/0141159031000076110} {\bibfield  {journal} {\bibinfo
   {journal} {Ph. Transit.}\ }\textbf {\bibinfo {volume} {76}},\ \bibinfo
  {pages} {155} (\bibinfo {year} {2003})}\BibitemShut {NoStop}%
\bibitem [{\citenamefont {Brüesch}\ and\ \citenamefont
  {Schüler}(1971)}]{Brueesch1971}%
  \BibitemOpen
  \bibfield  {author} {\bibinfo {author} {\bibfnamefont {P.}~\bibnamefont
  {Brüesch}}\ and\ \bibinfo {author} {\bibfnamefont {C.}~\bibnamefont
  {Schüler}},\ }\bibfield  {title} {\bibinfo {title} {Raman and infrared
  spectra of crystals with α-{NaFeO}$_2$ structure},\ }\href
  {https://doi.org/https://doi.org/10.1016/S0022-3697(71)80347-5} {\bibfield
  {journal} {\bibinfo  {journal} {J. Phys. Chem. Solids}\ }\textbf {\bibinfo
  {volume} {32}},\ \bibinfo {pages} {1025} (\bibinfo {year}
  {1971})}\BibitemShut {NoStop}%
\bibitem [{\citenamefont {Yelisseyev}\ \emph {et~al.}(2015)\citenamefont
  {Yelisseyev}, \citenamefont {Krinitsin}, \citenamefont {Isaenko},\ and\
  \citenamefont {Grazhdannikov}}]{Yelisseyev2015}%
  \BibitemOpen
  \bibfield  {author} {\bibinfo {author} {\bibfnamefont {A.}~\bibnamefont
  {Yelisseyev}}, \bibinfo {author} {\bibfnamefont {P.}~\bibnamefont
  {Krinitsin}}, \bibinfo {author} {\bibfnamefont {L.}~\bibnamefont {Isaenko}},\
  and\ \bibinfo {author} {\bibfnamefont {S.}~\bibnamefont {Grazhdannikov}},\
  }\bibfield  {title} {\bibinfo {title} {Spectroscopic properties of nonlinear
  optical {LiGaTe}$_2$ crystal},\ }\href
  {https://doi.org/https://doi.org/10.1016/j.optmat.2014.12.046} {\bibfield
  {journal} {\bibinfo  {journal} {Opt. Mater.}\ }\textbf {\bibinfo {volume}
  {42}},\ \bibinfo {pages} {276} (\bibinfo {year} {2015})}\BibitemShut
  {NoStop}%
\bibitem [{\citenamefont {Hoang}\ and\ \citenamefont
  {Mahanti}(2016)}]{Hoang2016}%
  \BibitemOpen
  \bibfield  {author} {\bibinfo {author} {\bibfnamefont {K.}~\bibnamefont
  {Hoang}}\ and\ \bibinfo {author} {\bibfnamefont {S.~D.}\ \bibnamefont
  {Mahanti}},\ }\bibfield  {title} {\bibinfo {title} {Atomic and electronic
  structures of {I-V-VI}$_2$ ternary chalcogenides},\ }\href
  {https://doi.org/https://doi.org/10.1016/j.jsamd.2016.04.004} {\bibfield
  {journal} {\bibinfo  {journal} {J. Sci. Adv. Mater. Devices}\ }\textbf
  {\bibinfo {volume} {1}},\ \bibinfo {pages} {51} (\bibinfo {year}
  {2016})}\BibitemShut {NoStop}%
\bibitem [{\citenamefont {Julien}(2003)}]{Julien2003}%
  \BibitemOpen
  \bibfield  {author} {\bibinfo {author} {\bibfnamefont {C.}~\bibnamefont
  {Julien}},\ }\bibfield  {title} {\bibinfo {title} {Lithium intercalated
  compounds: Charge transfer and related properties},\ }\href
  {https://doi.org/https://doi.org/10.1016/S0927-796X(02)00104-3} {\bibfield
  {journal} {\bibinfo  {journal} {Materials Science and Engineering: R:
  Reports}\ }\textbf {\bibinfo {volume} {40}},\ \bibinfo {pages} {47} (\bibinfo
  {year} {2003})}\BibitemShut {NoStop}%
\bibitem [{\citenamefont {Morris}(2009)}]{Morris2009}%
  \BibitemOpen
  \bibfield  {author} {\bibinfo {author} {\bibfnamefont {G.~A.}\ \bibnamefont
  {Morris}},\ }\bibfield  {title} {\bibinfo {title} {Spin dynamics: Basics of
  nuclear magnetic resonance, second edition. malcolm h. levitt. wiley
  chichester. 2008. pp xxv + 714. isbn 978-0-470-51118-3(hbk) 978-0-470-51117-6
  (pbk)},\ }\href {https://doi.org/https://doi.org/10.1002/nbm.1356} {\bibfield
   {journal} {\bibinfo  {journal} {NMR in Biomedicine}\ }\textbf {\bibinfo
  {volume} {22}},\ \bibinfo {pages} {240} (\bibinfo {year} {2009})}\BibitemShut
  {NoStop}%
\bibitem [{\citenamefont {Freude}\ and\ \citenamefont
  {Haase}(1993)}]{Freude1993}%
  \BibitemOpen
  \bibfield  {author} {\bibinfo {author} {\bibfnamefont {D.}~\bibnamefont
  {Freude}}\ and\ \bibinfo {author} {\bibfnamefont {J.}~\bibnamefont {Haase}},\
  }\bibinfo {title} {Quadrupole effects in solid-state nuclear magnetic
  resonance}\ (\bibinfo  {publisher} {Springer},\ \bibinfo {year} {1993})\ pp.\
  \bibinfo {pages} {1--90}\BibitemShut {NoStop}%
\bibitem [{\citenamefont {Ko}\ \emph {et~al.}(2014)\citenamefont {Ko},
  \citenamefont {Oh}, \citenamefont {Lee}, \citenamefont {Park}, \citenamefont
  {Kim},\ and\ \citenamefont {Choi}}]{ko2014}%
  \BibitemOpen
  \bibfield  {author} {\bibinfo {author} {\bibfnamefont {Y.-H.}\ \bibnamefont
  {Ko}}, \bibinfo {author} {\bibfnamefont {M.-W.}\ \bibnamefont {Oh}}, \bibinfo
  {author} {\bibfnamefont {J.~K.}\ \bibnamefont {Lee}}, \bibinfo {author}
  {\bibfnamefont {S.-D.}\ \bibnamefont {Park}}, \bibinfo {author}
  {\bibfnamefont {K.-J.}\ \bibnamefont {Kim}},\ and\ \bibinfo {author}
  {\bibfnamefont {Y.-S.}\ \bibnamefont {Choi}},\ }\bibfield  {title} {\bibinfo
  {title} {Structural studies of {AgSbTe}$_2$ under pressure: Experimental and
  theoretical analyses},\ }\href@noop {} {\bibfield  {journal} {\bibinfo
  {journal} {Curr. Appl. Phys.}\ }\textbf {\bibinfo {volume} {14}},\ \bibinfo
  {pages} {1538} (\bibinfo {year} {2014})}\BibitemShut {NoStop}%
\bibitem [{\citenamefont {Wojciechowski}\ \emph {et~al.}(2010)\citenamefont
  {Wojciechowski}, \citenamefont {Schmidt}, \citenamefont {Tobola},
  \citenamefont {Koza}, \citenamefont {Olech},\ and\ \citenamefont
  {Zybała}}]{Wojciechowski2010}%
  \BibitemOpen
  \bibfield  {author} {\bibinfo {author} {\bibfnamefont {K.}~\bibnamefont
  {Wojciechowski}}, \bibinfo {author} {\bibfnamefont {M.}~\bibnamefont
  {Schmidt}}, \bibinfo {author} {\bibfnamefont {J.}~\bibnamefont {Tobola}},
  \bibinfo {author} {\bibfnamefont {M.}~\bibnamefont {Koza}}, \bibinfo {author}
  {\bibfnamefont {A.}~\bibnamefont {Olech}},\ and\ \bibinfo {author}
  {\bibfnamefont {R.}~\bibnamefont {Zybała}},\ }\bibfield  {title} {\bibinfo
  {title} {Influence of doping on structural and thermoelectric properties of
  {AgSbSe}$_2$},\ }\href {https://doi.org/10.1007/s11664-009-1008-8} {\bibfield
   {journal} {\bibinfo  {journal} {J. Electron. Mater.}\ }\textbf {\bibinfo
  {volume} {39}},\ \bibinfo {pages} {2053} (\bibinfo {year}
  {2010})}\BibitemShut {NoStop}%
\bibitem [{\citenamefont {Milenov}\ \emph {et~al.}(2014)\citenamefont
  {Milenov}, \citenamefont {Tenev}, \citenamefont {Miloushev}, \citenamefont
  {Avdeev}, \citenamefont {Luo},\ and\ \citenamefont {Chou}}]{milenov2014}%
  \BibitemOpen
  \bibfield  {author} {\bibinfo {author} {\bibfnamefont {T.}~\bibnamefont
  {Milenov}}, \bibinfo {author} {\bibfnamefont {T.}~\bibnamefont {Tenev}},
  \bibinfo {author} {\bibfnamefont {I.}~\bibnamefont {Miloushev}}, \bibinfo
  {author} {\bibfnamefont {G.}~\bibnamefont {Avdeev}}, \bibinfo {author}
  {\bibfnamefont {C.-W.}\ \bibnamefont {Luo}},\ and\ \bibinfo {author}
  {\bibfnamefont {W.-C.}\ \bibnamefont {Chou}},\ }\bibfield  {title} {\bibinfo
  {title} {Preliminary studies of the raman spectra of {Ag}$_2${Te} and
  {Ag}$_5${Te}$_3$},\ }\href@noop {} {\bibfield  {journal} {\bibinfo  {journal}
  {Opt. Quant. Electron.}\ }\textbf {\bibinfo {volume} {46}},\ \bibinfo {pages}
  {573} (\bibinfo {year} {2014})}\BibitemShut {NoStop}%
\bibitem [{\citenamefont {Pandiaraman}\ \emph {et~al.}(2011)\citenamefont
  {Pandiaraman}, \citenamefont {Soundararajan},\ and\ \citenamefont
  {Ganesan}}]{pandiaraman2011}%
  \BibitemOpen
  \bibfield  {author} {\bibinfo {author} {\bibfnamefont {M.}~\bibnamefont
  {Pandiaraman}}, \bibinfo {author} {\bibfnamefont {N.}~\bibnamefont
  {Soundararajan}},\ and\ \bibinfo {author} {\bibfnamefont {R.}~\bibnamefont
  {Ganesan}},\ }\bibfield  {title} {\bibinfo {title} {Optical studies of
  physically deposited nano-{Ag}$_2${Te} thin films},\ }in\ \href@noop {}
  {\emph {\bibinfo {booktitle} {Defect and Diffusion Forum}}},\ Vol.\ \bibinfo
  {volume} {319}\ (\bibinfo {organization} {Trans Tech Publ},\ \bibinfo {year}
  {2011})\ pp.\ \bibinfo {pages} {185--192}\BibitemShut {NoStop}%
\bibitem [{\citenamefont {Sun}\ \emph {et~al.}(2023)\citenamefont {Sun},
  \citenamefont {Hong}, \citenamefont {He}, \citenamefont {Jian}, \citenamefont
  {Ju}, \citenamefont {Cai},\ and\ \citenamefont {Liu}}]{sun2023}%
  \BibitemOpen
  \bibfield  {author} {\bibinfo {author} {\bibfnamefont {F.}~\bibnamefont
  {Sun}}, \bibinfo {author} {\bibfnamefont {W.}~\bibnamefont {Hong}}, \bibinfo
  {author} {\bibfnamefont {X.}~\bibnamefont {He}}, \bibinfo {author}
  {\bibfnamefont {C.}~\bibnamefont {Jian}}, \bibinfo {author} {\bibfnamefont
  {Q.}~\bibnamefont {Ju}}, \bibinfo {author} {\bibfnamefont {Q.}~\bibnamefont
  {Cai}},\ and\ \bibinfo {author} {\bibfnamefont {W.}~\bibnamefont {Liu}},\
  }\bibfield  {title} {\bibinfo {title} {Synthesis of ultrathin topological
  insulator $\beta$-{Ag}$_2${Te} and {Ag}$_2${Te/WSe}$_2$-based
  high-performance photodetector},\ }\href@noop {} {\bibfield  {journal}
  {\bibinfo  {journal} {Small}\ }\textbf {\bibinfo {volume} {19}},\ \bibinfo
  {pages} {2205353} (\bibinfo {year} {2023})}\BibitemShut {NoStop}%
\end{thebibliography}%

\clearpage
\onecolumngrid

\large\textbf{Supplemental Material for \textquotedblleft On the ambient conditions crystal structure of AgSbTe$_2$ \textquotedblright}

\renewcommand{\thefigure}{S\arabic{figure}}
\setcounter{figure}{0}
\renewcommand{\thetable}{S\Roman{table}}

\begin{figure}[H]
\centering
\includegraphics[width=\linewidth]{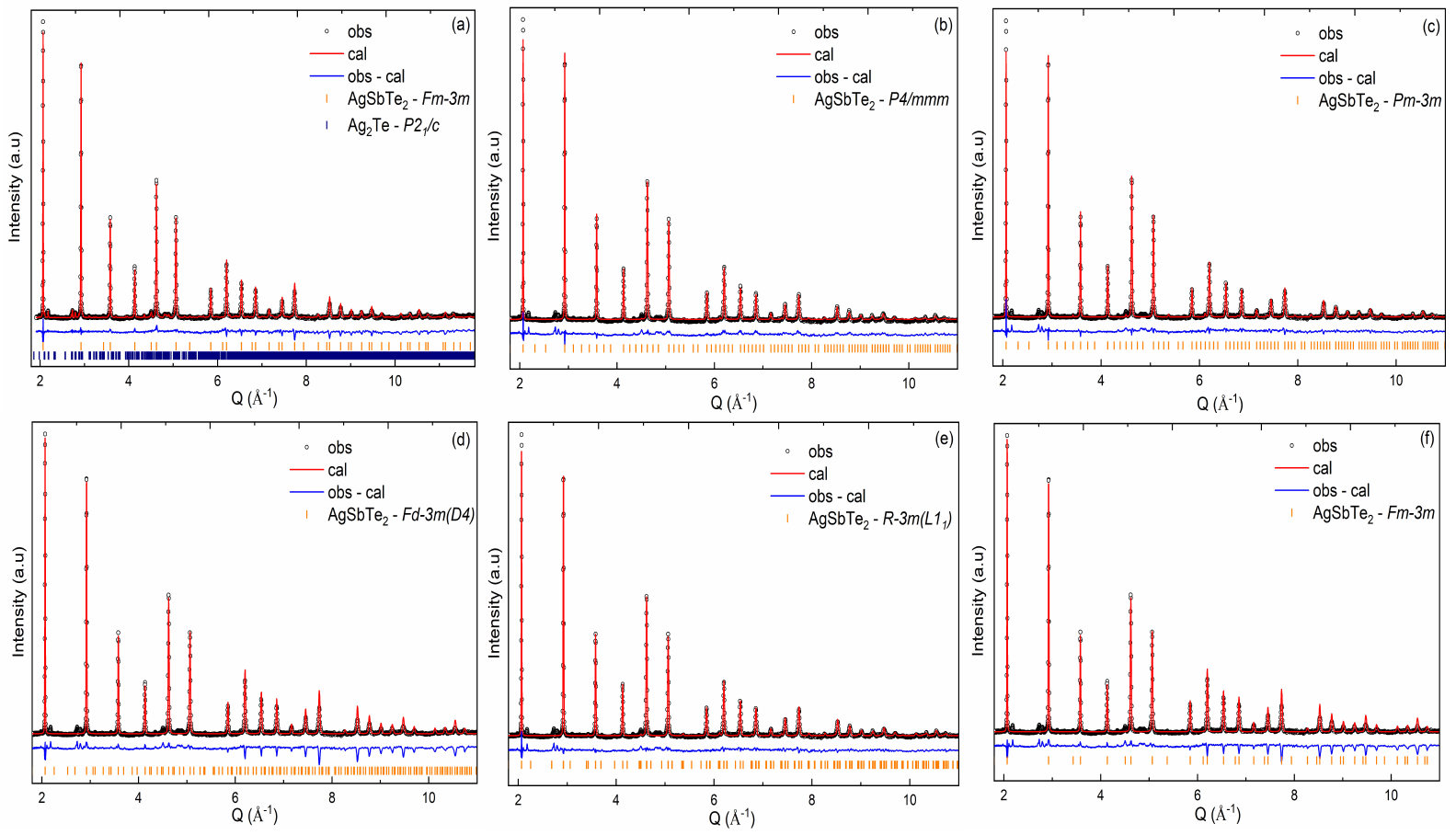}
\caption{Rietveld refinements of the experimental Neutron powder diffraction pattern of AgSbTe$_2$ using: a) the  \textit{Fm-3m} structure and the $\beta$-Ag$_2$Te (\textit{P2$_1$/c})  as secondary phase, (b) the\textit{ P4/mmm}, (c) the \textit{Pm-3m}, (d) the \textit{Fd-3m}, (e) the \textit{R-3m} and (f) the \textit{Fm-3}m structures as structural models.}
\end{figure}

\begin{figure}[H]
\centering
\includegraphics[width=0.7\linewidth]{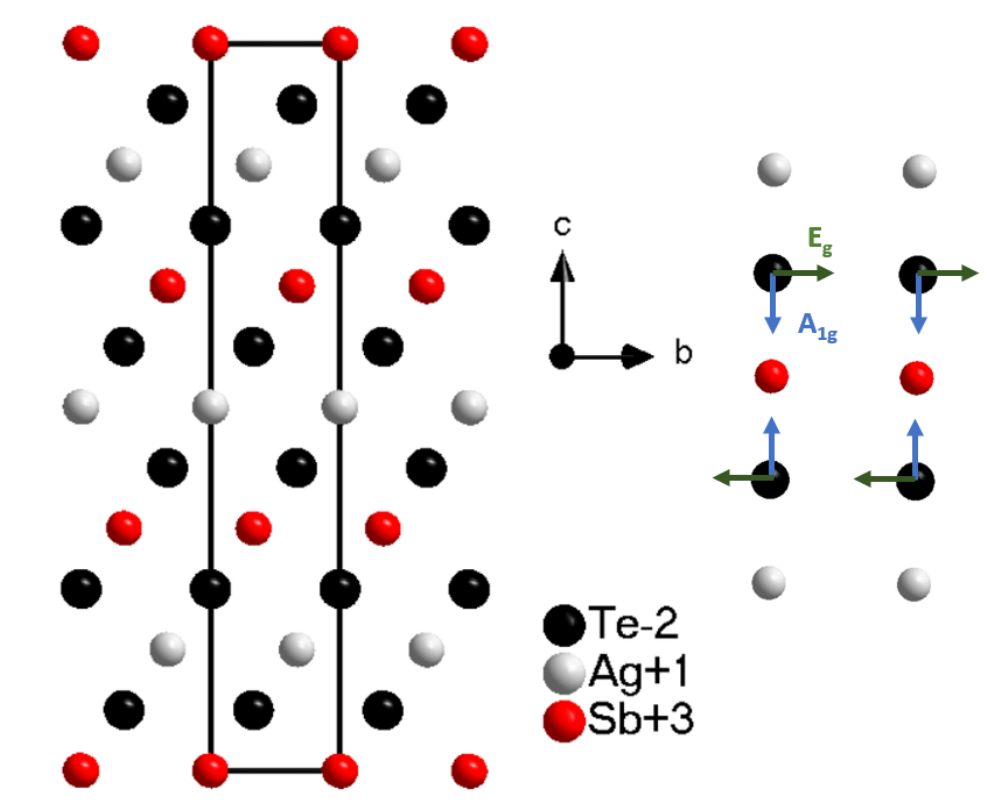}
\caption{Right, $\alpha$-NaFeO$_2$-type crystal structure of AgSbTe$_2$ and left, Raman modes eigenvectors.}
\end{figure}

\end{document}